\documentclass[12pt]{article}

\usepackage[margin=3cm]{geometry}
\usepackage[latin1]{inputenc}
\usepackage{amsmath}
\usepackage{amsfonts}
\usepackage{amssymb}
\usepackage{mathrsfs}
\usepackage{fontenc}
\usepackage{graphicx}
\usepackage{authblk}

\newcommand{\la}{\langle}
\newcommand{\ra}{\rangle}
\newcommand{\beq}{\begin{equation}}
\newcommand{\eeq}{\end{equation}}

\begin{document}

\title{Quantum Correlations in Multipartite Quantum Systems}

\author{Thiago R. de Oliveira}

\affil{Instituto de F\'{\i}sica, Universidade Federal Fluminense, Av. Gal. Milton Tavares de Souza s/n, Gragoat\'a, 24210-346, Niter\'oi, RJ, Brazil.}

\date{}
\maketitle

\begin{abstract}
We review some concepts and properties of quantum
correlations, in particular multipartite measures, geometric
measures and monogamy relations. We also discuss the relation between
classical and total correlations.
\end{abstract}

\section{Introduction}

Entanglement is usually said to be the characteristic trait of
quantum mechanics. All started with the recognition by Einstein, Podolsky
and Rosen \cite{Einstein35} that two-qubit states ssuch as the superposition
\beq
|\psi\ra=|00\ra + |11\ra,
\eeq
where $|0\ra$ and $|1\ra$ are the eigenstates of $\sigma^z$, have some kind of non-local "action at a distance" since a measure of
the first qubit somehow "changes" the state of the second qubit, no
matter how far away it is: if I measure one qubit and obtain $|0\ra$ ($|1\ra$),
I know immediately that a measurement on the other, in the same basis, will also return
the state $|0\ra$ ($|1\ra$). However it was later realized that such
states alone do not allow communication at a distance and therefore do not
violate the principle of special relativity. But such states do allow
for stronger correlations than allowed by a classical theory, as seen
in the violation of Bell inequalities. One way to see such
stronger correlations is to note that the perfect correlations between
measurements of the spin, are not true only for measurements along the $z$ direction, but actually
in any direction. As far as one deals with pure states, the situation is
clear. However, the generalization of the concept of entanglement to mixed
states is more complicated. Werner in 1989 \cite{Werner89}
proposed non-entangled, or separable mixed states, to be the ones
which can be written as
\beq
\rho=\sum_i p_i \;\rho_A^i \otimes \rho_B^i.
\eeq
This definition is motivated by the fact that these are the states
which can be created by two separated labs using local quantum
operations and classical communication: they contain only classical
correlations due to the classical communication. Entanglement was then
rigorously defined as a property of quantum states which can
not be created by local operation and classical communication (LOCC).
This framework of LOCC created the basis for entanglement theory which defines
good entanglement measures as the ones which do not increase under
LOCC. But already Werner \cite{Werner89} noted that such definition allowed for
entangled states which do not violated any Bell-type inequality,
opening a gap between the concept of entanglement and non-locality.
\footnote{There is a vast literature studying this gap and looking
for more general types of Bell inequality which may close the gap; see \cite{Brunner14}.}

And in 2002, studying the correlation between apparatus and system
in a measurement, Ollivier and Zurek realized \cite{Ollivier02}
that separable states, as defined by Werner, may still have some
quantumness in the sense that they can be perturbed by local measurements. Let's focus on "perfect" von Neumman measurements, defined by a set of one-dimensional orthogonal projectors $\{ \Pi_j^B \}$ on system $B$, the apparatus. The state of  $A$ after the outcome corresponding to 
$\Pi_j^B$ has been detected is
\beq
\rho_{A|\Pi_j^B} = \frac{ \Pi_j^B \rho_{AB} \Pi_j^B }{ \text{Tr}[\Pi_j^B \rho_{AB}]},
\eeq
and this outcome happens with probability $p_j=\text{Tr}[\Pi_j^B \rho_{AB}]$. 
It can be shown that the
only way that the state of $A$ is not perturbed by this measurement is if it 
can be written as
\beq
\chi_{AB}=\sum_i \rho_A^i \otimes \Pi_i^B.
\eeq
This is a separable state with only fully-distinguishable states (orthogonal ones)
for $B$ and some indistinguishable states for  $A$ . Such states are
called quantum-classical since there are measurements on $B$ which do not
perturb the state; but measurements on $A$ may perturb it.
States which can not be written in such form are perturbed by
all local measurements on $B$. This perturbation of the state of $A$
by local measurements on $B$ is a quantum aspect of the correlations
between $A$ and $B$ that goes beyond entanglement and that can be
quantified in many ways. The perturbation
will also in general decrease the correlation between the parts.

There are many possibilities to quantify this quantum aspect of the correlation. Lets consider the conditional entropy of $A$ given $B$:
\beq
S(\rho_A|\rho_B) = S(\rho_{AB})-S(\rho_B).
\eeq
Considering that entropy measures the uncertainty about the system, the
conditional entropy is the remaining uncertainty about $AB$ after we
learn the state of $B$ and it is associated with our uncertainty, on average, about A, given that we know the state of $B$; we measured it.
But as we mentioned, the acting of measuring the system can perturb it and thus  change the conditional  entropy. Therefore for quantum states we can define the conditional entropy in a alternative way as
\beq
S(\rho_{AB}|\Pi^B) = \sum_j p_j \; S(\rho_{A|\Pi_j^B}).
\eeq
It still has the interpretation of the average uncertainty about
A given that we measured $B$. Classically these two definitions are equivalent, but for some quantum states they can differ
\footnote{Actually $S(\rho_A|\rho_B)$ is always positive in classical setting, but
can be negative for entangled states and took it a long time to understand this negativity; see \cite{Horodecki05}. }.
 This difference in the
definition also propagates for other entropy measures of correlation.
The mutual information, for example, can be written as
\beq
I(\rho_{AB})=S(\rho_A)+S(\rho_B)-S(\rho_{AB})
\eeq
or in terms of the conditional entropy
\beq
J_{\Pi^B}(\rho_{AB}) = S(\rho_A)-S(\rho_{AB}|\Pi^B),
\eeq
where we used the alternative definition for the conditional entropy and
another symbol, $J$, since the two definitions of the mutual information
may not be equivalent. The first expression suggests the interpretation
of the mutual information as the common information between $A$ and $B$ and
therefore as the measure of its total correlation. The second expression
suggests that the mutual information is a measure of the decrease on
uncertainty, or gain in information, about $A$ as a result of a measurement on $B$. The second definition was introduced in
\cite{Henderson01} as a measure of the classical correlation.

Based on these considerations the discord was defined by Ollivier and Zurek \cite{Ollivier02}
as 
\beq
D_{\Pi^B}(\rho_{AB}) = I(\rho_{AB}) - J_{\Pi^B}(\rho_{AB}).
\eeq
It measures how much common information, or correlation, was lost
in the measurement. In other words, it measures the information about
$A$ that exists in the correlation but can not be extracted locally by 
reading the state of $B$.  It can also be written as the difference between 
the two definitions of conditional entropy: 
$S(\rho_A|\rho_B)-S(\rho_{AB}|\Pi^B)$.
One should minimize over all possible measurements
on $B$ to find the one which disturbs the least $A$ and allows us to extract 
the most
information about $A$ by measuring B. Thus the measurement independent discord
was defined as
\beq
D_B(\rho_{AB}) =\min_{\Pi^B} D_{\Pi^B}(\rho_{AB}).
\eeq
The discord has the following properties: i) it is not symmetric
under the change of $A$ for B; ii) it is non-negative; iii) it vanishes
if and only if the state is quantum-classical; iv) it is invariant
under local unitary transformations. Unfortunately it does
not have an important property for correlations measures: to
not increase under local operations
\footnote{We should stress that it is a natural requirement that correlation
measurements should not increase under local operations: one should no be able
to increase their correlation with someone far away only acting on their own system.}. It may increase by simple local operations and therefore is not a bona
fine measure of correlations. In sum, discord does indicate that the correlation in the state has a quantum aspect, but it is not a bona fide quantifier of the amount of such correlation.

There are also many other possibilities to quantify this quantumness
in separable states. And in fact many discord-like measures were proposed,
and still are being proposed (see \cite{Modi12} for a review). We will just mention
another one since it will be of our interest later, and actually is closely related to the original definition of discord. The idea is to consider
local von Neumann measurements on both parties. The state after the
non-selective measurement is
\beq
\Pi^{AB}(\rho_{AB})= \sum_{i,j} (\Pi_i^A \otimes \Pi_j^B) \rho_{AB}  (\Pi_i^A \otimes \Pi_j^B)
\eeq
and has the general form
\beq
\chi_{AB} = \sum_{i,j} p_{ij} \; \Pi^A_i \otimes \Pi^B_j.
\eeq
Such states are called classical-classical, since they are
the ones which are not perturbed by the local measurement on $A$ or $B$.
Thus the probability $p_{ij}$ can be regarded as a classical joint
probability of the random variables $i$ and $j$. One then defines
the symmetric discord as
\beq
D_S(A:B)= \min_{\Pi^{AB}} \;[ I(\rho_{AB}) - I(\Pi^{AB}(\rho_{AB})) ].
\eeq
This idea was originally proposed as a measurement-induced disturbance
and without the optimization over measurements. It was then redefined
with the optimization  and studied by several authors (see Sec. II.E of
\cite{Modi12}). Note also that it may be argued that the asymmetric 
discord is not a good measure of the quantum correlation since it could be null from one side and not the other. In other words, a quantum-classical state still has some quantumness 
in its correlations when measurements are
made on part A. It was also realized that the original asymmetric
discord is equivalent to the difference between the mutual
information before and after a local measurement is made on one of the parts. 

For the sake of completeness we should mention that a natural
question is what would happen if one considers positive operator
valued measurement (POVM), which are more general than von Neumann
projective measures. And the question is if the minimum
discord is attained with von Neumann measurements, in which
case considering POVM would not be necessary. It can be
shown that von Neumann are not always optimum but extremal
rank-one POVM are sufficient (see Sec. II.I of \cite{Modi12}).

\section{Multipartite quantum discord}

One important and difficult question when dealing with correlations is
how to extend them beyond the two-part scenario. In this multipartite
scenario there is not even a single conceptual framework, not to mention
measures. One can for example consider many different bipartitions
of the multipartite system. For three qubits we
could consider the correlation between one of the particles
and the rest. We could then average over all possible bipartitions
of one with the other two. Or we could take the minimum, or the maximum.
Actually we could consider any function of the possible combinations.
And for more than three particles there are even more options, since
besides the bipartition of one with the rest, we could still have
two with the rest, three with the rest and so on. Thus we have
many possible bipartitions and can still combine the correlations
in many different ways. Another possibility  would be not to
use a single number but build a correlation
vector (or matrix) to characterize the multipartite correlation.
It is clear then that the problem is very complex, something already
realized in the quantification of multipartite entanglement where
a zoo of measures exist, but still very little is well understood.

For discord we also have many possibilities. We could consider
measures only on single particles or in groups of particles.
This is equivalent to the many possible ways to write the
mutual information in terms of conditional entropies. For three particles the mutual information can be written as
\beq
I(\rho_{ABC})=S(\rho_A)+S(\rho_B)+S(\rho_C).
-S(\rho_{ABC})
\eeq
But in terms of the conditional entropy there are
many possible combinations. These would be the classical
correlation and two possibilities are
\beq
S(\rho_{AB}) - S(\rho_B|\rho_A) - S(\rho_A|\rho_B)
- S(\rho_A|\rho_C) - S(\rho_B|\rho_C) + S(\rho_{AB}|\rho_C)
\eeq
and
\beq
S(\rho_{A}) + S(\rho_B) + S(\rho_C)
- S(\rho_{AB}) - S(\rho_{AC}) + S(\rho_{A}|\rho_{BC})
\eeq
Note that the first case involves only single-particle measures, while the second one involves only two-particle
measures. One of the first works on multipartite discord proposed to use the two expressions above to define multipartite discord measures, which they called quantum
dissenssion for one- and two-particle measures \cite{Chakrabarty11}. And of course one could also combine all these quantities in many ways to define another multipartite measure or construct a correlation vector as proposed
in \cite{Sazim16}.

One proposal, which gained some attention \cite{Giorgi11}, started by defining a
symmetric version of discord as the minimum between the asymmetric discord in relation to $A$ and in relation to $B$: $D_{AB}=\min\{D_A,D_B\}$.
In the same way one can define the symmetric classical correlation as the maximum between the asymmetric ones:
$J_{AB}=\max\{J_A,J_B\}$. 
It then considers that the correlation in a tripartite system
can be decomposed in a bipartite part and a genuine tripartite part. This division should be true for the
total, classical and quantum correlation. The next step is to use conditional entropies involving both single and two particles measures as
\beq
J_{BC,B}(\rho_{ABC})=S(\rho_A) + S(\rho_B) - S(\rho_A|\rho_{BC})
- S(\rho_C|\rho_{B})
\eeq 
to define the total classical correlation (we are assuming a maximization over
all possible measures on $BC$ and $B$). Actually, there
are six possible definitions similar to the above with difference only on
the single or two parties being measured. And the total classical correlation is defined as the maximum among them:
$J(\rho_{ABC})=\max_{i,j,k} \{ J_{ij,k}(\rho_{ABC}) \} $. The bipartite part of this classical correlation is defined as $J^{(2)}=\max\{J_{AB},J_{AC},J_{BC} \}$. The genuine tripartite classical correlation is then the difference between the total and the bipartite classical
correlation: $J^{(3)}(\rho_{ABC})=J(\rho_{ABC})-J^{(2)}(\rho_{ABC})$. In the same way we can define the total, bipartite and genuine tripartite correlations. While the definitions may seem arbitrary they are interesting as they have nice properties and are related to the relative entropy (see \cite{Giorgi11} for more details).

There is also the option to consider sequential single-particle measures. In the bipartite case, one first makes the optimal measurement $\tilde{\Pi}^B$ on $B$ to get the discord $D_{\tilde{\Pi}^B}(\rho_{AB})$. One then makes the optimal measurement on part 
$A$ of $\tilde{\Pi}^B(\rho_{AB})$. We thus have a symmetric discord as the sum:
\beq
D_{\tilde{\Pi}^B}(\rho_{AB}) + D_{\tilde{\Pi}^A}(\tilde{\Pi}^B(\rho_{AB})).
\eeq
The generalization to the $N$-partite system is straightforward, one realizes the sequential optimal measurements on each particle adding the corresponding discords (see \cite{Okrasa11} for more details).

A natural generalization of discord for multipartite systems is
to extend the symmetric discord as defined by the mutual information
before and after a local measurement on both parts $\Pi^A\otimes\Pi^B$. This was named
global quantum discord and defined as \cite{Rulli11}
\beq
D(\rho_{AB})=\min_{\Pi^{A} \otimes \Pi^{B}}
\;\; [I(\rho_{AB}) - I(\Pi^{A}\otimes \Pi^{B}(\rho_{AB}))].
\eeq
Note that here measures can be done all together or sequentially
since they are local and commute. But one does not add the partial discords
after each measurement. For an arbitrary multipartite
state of $N$ parts the natural generalization is
\beq
D(\rho_{A_1...A_N})=\min_{\Pi^{A_1}\otimes...\otimes \Pi^{A_N}}
\;\; [I(\rho_{A_1...A_N}) - I(\Pi^{A_1}\otimes...\otimes \Pi^{A_N}(\rho_{A_1...A_N}))]
\eeq
It can be shown that this measure is non-negative.
It takes value one for the tripartite GHZ state. And when one considers
a mixture of the tripartite GHZ state with a fully mixed state
(the identity), one can show that the global discord decreases
with the decrease in the weight of the GHZ in the mixture, becoming null only for a zero contribution of the GHZ. 

It is also possible to define multipartite geometric measures of
discord. One just generalizes the definition of the set of product
and classical states to states of the form $\pi_1 \otimes ... \otimes \pi_N $
and $\sum p_{i_1...i_N} \pi_1 \otimes ... \otimes \pi_N $. Then we just
chooses a distance measure to define the discord, and even other
correlations. These measures may be viewed as true multipartite measures,
since one does not appeal to the use of many bipartitions.

\section{Geometric correlations}

As we mentioned before, one can use different figures of merit to quantify 
Discord. Most of these measures are related to entropy measures. A different\footnote{Note that the relative entropy can also be understood as a distance measure, even though technically it is not a genuine distance since it is not symmetric.} approach is the geometric one: to use the distance of a given state to the set of classical states (see fig. 2). Mathematically the geometric
Discord of a given state $\rho$ is given by
\beq
D_G(\rho) = \min_{\chi \in \mathcal{C}} ||\rho - \chi||^2
\eeq
where $||X||$ is a operator distance in the Hilbert space and $\mathcal{C}$
is the set of classical states (the ones with zero discord),
which are mixtures of locally distinguishables states
\beq
\chi = \sum_{i,j} p_{ij} \; \Pi^A_i \otimes \Pi^B_j,
\eeq
with $p_{ij}$ a joint probability distribution, $\Pi_i=|k_i\ra\la k_i|$ with local
states $|k_i\ra$ spanning a local orthonormal basis. Here one also has in principle many possible geometric measures using different distance measures.
These measures have the appeal of a geometric interpretation and for some
choices of distances can be interpreted as the distinguishability between states.

The first proposal of a geometric measure appeared in 2010 and used
the Hilbert-Schmidt norm \cite{Dakic10}. The Hilbert-Schmidt norm,
also known as the 2-norm, of an operator $X$ is given by
\beq
||X||^ 2_2=\text{Tr}(X X^\dagger).
\eeq
This norm is the most used one since it is easy to evaluate.
It thus allows for a closed expression of the geometrical discord of two qubits
(see \cite{Dakic10}), what is an important feature since there are no analytical expressions for most entropic discord measures. As it was easy to calculate the
geometric discord using the Hilber-Schmidt norm was used in many works, including 
experimental papers, and it was even related to the performance of remote state preparation \cite{Dakic12,Tufarelli12}. Nonetheless, it was found that it 
was not a proper measure of quantumness of correlations, since it can
increase under trivial local reversible operations on the unmeasured party \cite{Piani12}. To see this, consider a simple map (channel) $\Gamma^\sigma:X \rightarrow X \otimes\sigma$: it introduces an ancilla which can be noisy but is uncorrelated. Under such operation we have
\beq
||X||_2 \rightarrow ||X\otimes\sigma||_2 = ||X||_2 ||\sigma||_2 =
||X||_2 \sqrt{\text{Tr}(\sigma^2)},
\eeq
where in the second equality we used that the Hilbert-Schmidt norm is multiplicative on tensor products. With this we have that
\beq
D_G(\Gamma^\sigma_B(\rho_{AB}))=D_G(\rho_{AB}) \text{Tr}[\sigma^2].
\eeq
Therefore just adding, or removing, a local, uncorrelated and noisy ancilla ($\text{Tr}[\sigma^2]<1$) in the unmeasured part $B$, a reversible operation, can change the Discord. But, if one remembers that Discord measures are not monotonic under local operations anyway, this can be considered not a fundamental problem.
But on the other side this is a very trivial local operation: we are just
adding or removing an uncorrelated ancilla, which can actually always be 
there, and in the unmeasured part. We should also note that the map increases the dimension of part B, but there are also examples showing that the geometric discord can increase even for local maps which preserve the dimension \cite{Fan13}.

The origin of the problem with the Hilbert-Schmidt geometric discord lies in 
the fact that the Hilbert-Schmidt norm can increase under completely positive 
trace-preserving (CPTP) maps; it can increase under quantum evolution. Actually this 
problem was already recognized after the proposals of geometric measures for entanglement. In the beginning of the development of entanglement theory, Vedral et al.
proposed three necessary conditions that any entanglement measure should satisfy.
\cite{Vedral97}.
They then showed that the distance between a state and the set of separable states
is a good entanglement measure (satisfying  their conditions) if the given distance
has the property of not increasing under CPTP maps: 
$D(\Gamma(\rho),\Gamma(\sigma))\leq D(\rho,\sigma)$. But later it was shown that
it was not the case for the Hilbert-Schmidt norm \cite{Ozawa00}.
Thus a possible solution is to use a contractive norm, as usually one
calls norms which do not increase under CPTP maps.

We should also mention  that there are other possibilities to fix the problem
raised in \cite{Piani12}. One could
redefine the measure taking the supremum over all maps on the unmeasured
part \cite{Piani12}. Besides seeming a bit artificial, this measure is in principle much
more difficult to calculate. Another possibility it to rescale the measure
by the state purity \cite{Tufarelli13}. However in both cases problems
are still expected to appear from the non-contractive property of the
Hilbert-Schmidt norm. 

Soon after the problem was raised, \cite{Paula13} proposed to use the 
trace norm for the geometric discord and obtained an analytical expression for a class of states: the Bell-diagonal states. Actually they considered an
general Schatten $p$-norm Discord. The Schatten norm
\footnote{There are many different ways to define a norm for a matrix (or operator).
One should first consider the p-norms of a vector $\vec{v}$ given by 
$||\vec{v}||=(\sum_i |v_i|^ p)^ {1/p}$ with $v_i$ being the components of $\vec{v}$
in some basis and $p\geq 1$. For $p=2$ we have the Euclidean norm.
One can then define the induced norm for the matrix as the maximum norm the
matrix can induce in a unit vector: $||X||=\sup_{|\vec{v}|=1} ||X\vec{v}||$.
Then given a vector p-norm vector we get a operator p-norm. Another possibility
is to consider an $m$ x $n$ matrix as a $mn$ vector and use an vector norm. These are
usually called "entrywise" norms. A third possibility, the Schatten norms, is to apply
the vector p-norm to the singular values of the matrix (the singular values
are the square root of the eingenvalues of $X^\dagger X$). For $p=2$ we have
the Hilbert-Schmidt, also called Frobenius, norm which is equivalent to the $p=2$
entrywise norm mentioned before. For $p=1$ we have the trace norm and for 
$p=\infty$ we have the spectral norm which is equivalent to the induced $p=2$
norm and also called operator norm and given by the largest singular value.
}
of an operator is given by 
$||X||_p=\text{Tr}[(X^\dagger X)^\frac{p}{2}]^\frac{1}{p}$
and they are multiplicative under tensor products:
$||\Gamma\sigma(X)||_p=||X||_p ||\sigma||_p$. We then define the p-Schatten
geometric Discord as
\beq
D_p(\rho) = \min_{\chi \in \mathcal{C}} ||\rho - \chi||^p_p.
\eeq
It is trivial to note that 
\beq
D_p(\Gamma^\sigma_B(\rho_{AB}))=D_p(\rho_{AB}) ||\sigma||_p^p.
\eeq
As density operators are Hermitian we have that $||\sigma||_p=\text{Tr}[\sigma^p]^{1/p}$. And as $\text{Tr}[\sigma]=1$ we
have that  $||\sigma||_p=1$ if and only if $p=1$. Therefore the only
$p$-Schatten geometric Discord which does not increase under the removal
or addition of local ancillas is the trace norm. Even more, as the trace
norm is contractive under the CPTP maps, the trace distance geometric
Discord can not increase under local operations in the unmeasured system.
This would also be the case for other contractive distance measures
as the Bures and Hellinger distances. 

The trace norm geometric Discord is then a bona fide measure of correlations. 
It is also related to the probability of distinguishing between two states 
via a single measurement
\footnote{This follows directly from the
fact that the trace distance itself has a interpretation in terms
of distinguishability: Suppose Alice prepares a quantum system in
state $\rho$ with probability 1/2 and in state $\sigma$ with 
probability 1/2. She then gives the system to Bob, who performs a 
POVM measurement to distinguish the two states. It can be shown that
Bob's probability of correctly identifying which state Alice prepared
is $1/2+1/2||\rho-\sigma||_1$ (see sec. 9.2 of \cite{Nielsen00}).}
and to another measure of quantumness
of correlation, the negativity of quantumness, when the measured
part is a qubit \cite{Nakano13}. In this measure the quantum correlation
is defined as the minimum entanglement created between a system and
a measurement apparatus by a local measurement. One drawback of the
trace distance discord is that is not as simple to calculate  as the
Hilbert-Schmidt norm. In fact there is still no closed analytical
expression for general two-qubit states, but only for some classes.
An expression for the Bell diagonal states was presented first
in \cite{Paula13}, but assuming that the closest classical
state also has the Bell diagonal form. Such assumption was
confirmed numerically for random states. Later, using a
different approach, the same formula was obtained without
any assumption \cite{Nakano13}; they also obtained a closed
expression for Werner, isotropic states and for all two-qubit states for 
which the reduced state of the measured systems is maximally mixed.
More recently, the optimization problem for general two-qubit states was shown to be equivalent to the minimization of a two-variable function (but which 
parametrically depends on the Bloch vectors of the reduced density matrix
and the singular values of the correlation matrix) and
a closed expression for X states was also obtained \cite{Ciccarello14}.
We should also mention that a general analytical expression exists
for the geometric discord using the Hellinger distance \cite{Roga16}.

It is also possible to define a measurement-induced geometric measure of
discord as \cite{Luo10}
\beq
D_{MG}(\rho) = \min_{\Pi^B} D(\rho-\Pi^B(\rho)),
\eeq
where the minimum is over all von Neumann measurements (rank one projectors)  $\Pi^B=\{ \Pi^B_i \}$ on the part $B$ and 
$\Pi^B(\rho)=\sum_i (\bf{1^A \otimes \Pi^B_i})\rho(\bf{1^A \otimes \Pi^B_i})$
is the post measurement state in the absence of readout. Here again there are many possibilities of distance measure to use. When using the same distance 
it is clear that $D_{MG}(\rho) \geq D_{G}(\rho)$ since in the measurement-induced
one needs to optimize only over classical states generated by von Neumann measurements.
The two measurements are equivalent when using the Hilbert-Schmidt distance.
For the trace distance the equivalence is true only if the system $A$ is
a qubit. They are generally different when using the Bures  and Hellinger
distances \cite{Roga16}.
	
\begin{figure}
\centering
\includegraphics[scale=0.4]{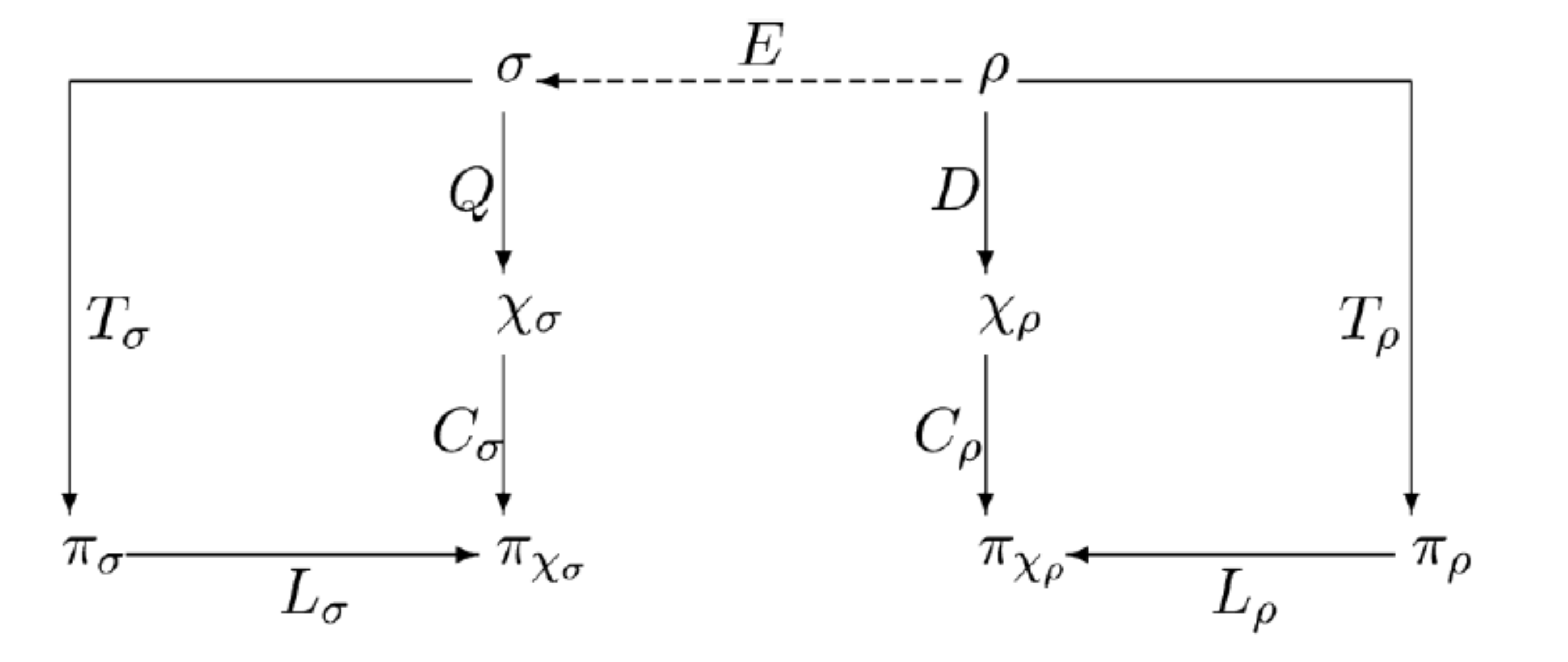}
\label{Fig:UnifiedView}
\caption{From \cite{Modi12}. The arrows represent the closest state using the relative entropy. $\rho$ is a entangled state, $\sigma$ a separable state, $\chi$ a classical state and $\pi$ a product state. $E$ is the entanglement, $D$ the Discord,
$Q$ the quantum dissonance, $T_\sigma$ and $T_\rho$ the total mutual information, $C_\sigma$ and $C_\rho$ the classical correlation. $L_\sigma$ and $L_\rho$ have no physical interpretation but allow for an additivity relation between the quantities: $T_\rho=D+C_\rho-L_\rho$ and analogously for $\sigma$.}
\end{figure}

One last possibility we should mention is to use the
relative entropy as a measure of distance. The quantum relative entropy, is defined as
\beq
S(\rho||\sigma)=\text{Tr}(\rho\log\rho - \rho\log\sigma).
\eeq
This entropy is also named the Kullback-Leibler divergence and
is often used to distinguish two probability distributions or density operators. It resembles a distance measure but
strictly it is not one since it is not symmetric. However, it
was originally proposed as a possible unified view on the
quantum and classical correlations. Besides the set of
separable and classical states one also defines the
set of product states,
\beq
\pi=\pi_1 \otimes...\otimes\pi_N
\eeq
as the states having no correlation at all. The separable
states are mixtures of general product states and classical
states are mixtures of product states which are orthogonal.
Then discord is defined as
\beq
D(\rho)=\min_{\chi \in \mathcal{C}} S(\rho||\chi)
\eeq
and entanglement as
\beq
E(\rho)=\min_{\sigma \in \mathcal{S}} S(\rho||\sigma).
\eeq
For an entangled state $\rho$, the discord $D(\rho)$ can
also contain some entanglement. So we can look at the
closest classical state to the closest separable state $\sigma$,
denoted by $\chi_\sigma$. This distance contains the non-classical
correlation excluding entanglement and was named quantum
dissonance (see fig. 1 and 2)
\beq
Q(\sigma) = \min_{\chi \in \mathcal{C}} S(\sigma||\chi)
\eeq
Further, we can compute the classical correlation
as the minimal distance between a classically correlated
state and a product state
\beq
C(\chi) = \min_{\pi \in \mathcal{P}} S(\chi||\pi)
\eeq
and the total correlations as the distance of the original
state to the closest product state
\beq
T(\rho) = \min_{\pi \in \mathcal{P}} S(\rho||\pi).
\eeq
These distances are illustrated in fig. \ref{Fig:UnifiedView}. Note that
for an entangled state $\rho$ one has both $T(\rho)$
and $T(\sigma)$, and that given a state $\rho$ one
can first find the closest classical state $\chi_\rho$ and then
look for the closest product state to $\chi_\rho$, which is
$\pi_{\chi_\rho}$. Or one can directly look at the closest product state to $\rho$, $\pi_\rho$. The two states are
not equal and we then define
\beq
L(\rho) =  S(\pi_\rho||\pi_{\chi_\rho})
\eeq
and similarly $L(\sigma)$. These two last quantities
allow additivity conditions for correlations as illustrated
in fig.\ref{Fig:UnifiedView}. 

The advantage of using the relative entropy is that
it is possible to find the closest product states and then
a closed expression for some of the correlation measures. 
First, it is easy to show (as done in \cite{Modi10}) that "the closest product
state to any state $\rho$, as measured by the relative entropy, is its reduced 
state in the product form, .i.e., $\pi_\rho=\pi_1\otimes...\otimes\pi_N$". 
Thus, using the linearity of trace and additivity of $\log$ we easily obtain
\begin{align}
T_\rho \equiv S(\rho||\pi_\rho) &=- \text{tr}[\rho\log(\pi_1\otimes...\otimes\pi_N) + \rho\log\rho] \\
& =\sum_i -\text{tr}(\pi_i\log\pi_i) + \text{tr}(\rho\log\rho) \\
& =S(\pi_\rho) - S(\rho).
\end{align}
This also allows us to write
\beq
C_\rho=S(\pi_{\chi_\rho}) - S(\chi_\rho)
\eeq
and
\beq
C_\sigma=S(\pi_{\chi_\sigma}) - S(\chi_\sigma).
\eeq
It is also possible to show that the closest classical state to a generic state $\rho$ is $\chi_\rho=\sum_{\vec{k}} \;
|\vec{k}\ra \la \vec{k}| \rho | \vec{k} \ra \la \vec{k}|$ with $\{|\vec{k}\ra\}$ the eigeinbasis of $\chi_\rho$. Then it is possible to show \cite{Modi10}
that the quantum correlations are also a difference between entropies
\beq
D = S(\chi_\rho) - S(\rho)
\eeq
\beq
Q = S(\chi_\sigma) - S(\sigma)
\eeq
with $S(\chi_\rho)= \min_{\vec{k}} S(\sum_{\vec{k}} |\vec{k}\ra \la \vec{k}| \rho | \vec{k} \ra \la \vec{k}|)$ and analogously to $S(\chi_\sigma)$. 
The minimization involved in $D$ and $Q$ is only over the possible local basis ${\vec{k}}$. However, this is equivalent to minimizing over all rank-one POVM measurements and is still a very difficult problem. Note that finding the
closest separable state to obtain the relative entropy of entanglement is 
also a hard task. Finally it is possible
to show that
\beq
L_\rho = S(\pi_{\chi_\rho}) - S(\pi_\rho)
\eeq
and
\beq
L_\sigma = S(\pi_{\chi_\sigma}) - S(\pi_\sigma).
\eeq
So all the quantities can be written as the difference of
entropies, the entropic cost, of performing operations that
bring the states to the closest one without the desired property;
it is the entropic cost of destroying the given correlation.
This simple relation also allows us to obtain the following
additivity inequalities:
\beq
T_\rho = D + C_\rho - L_\rho
\eeq
\beq
T_\sigma = Q + C_\sigma - L_\sigma 
\eeq
which justify the diagram of fig. \ref{Fig:UnifiedView}.
We should also mention a relation with the original entropic
discord. For this we should consider the set of classical-quantum
states instead of the classical-classical states. In this case
it is possible to show that the original entropic discord is
equal to $D-L_\rho$ or equivalently to $T_\rho - C_\rho$,
but only when considering a minimization over only projective
measurements for the original entropic discord.

It is also possible to define the geometric total and classical correlation
using other distance measures. This was studied for the Trace distance in \cite{Paula13B,Paula14}, for the Hilbert-Schmidt norm in
\cite{Bellomo12} and for the Bures distance in \cite{Bromley12},
but the separation of the total correlation in
a classical and quantum part does not obey the additivity relation anymore.
Even the first property that the closest product state is the product of 
the marginals is not true in general, which makes the computation of
the total correlation non-trivial.

\begin{figure}
\centering
\includegraphics[scale=0.2]{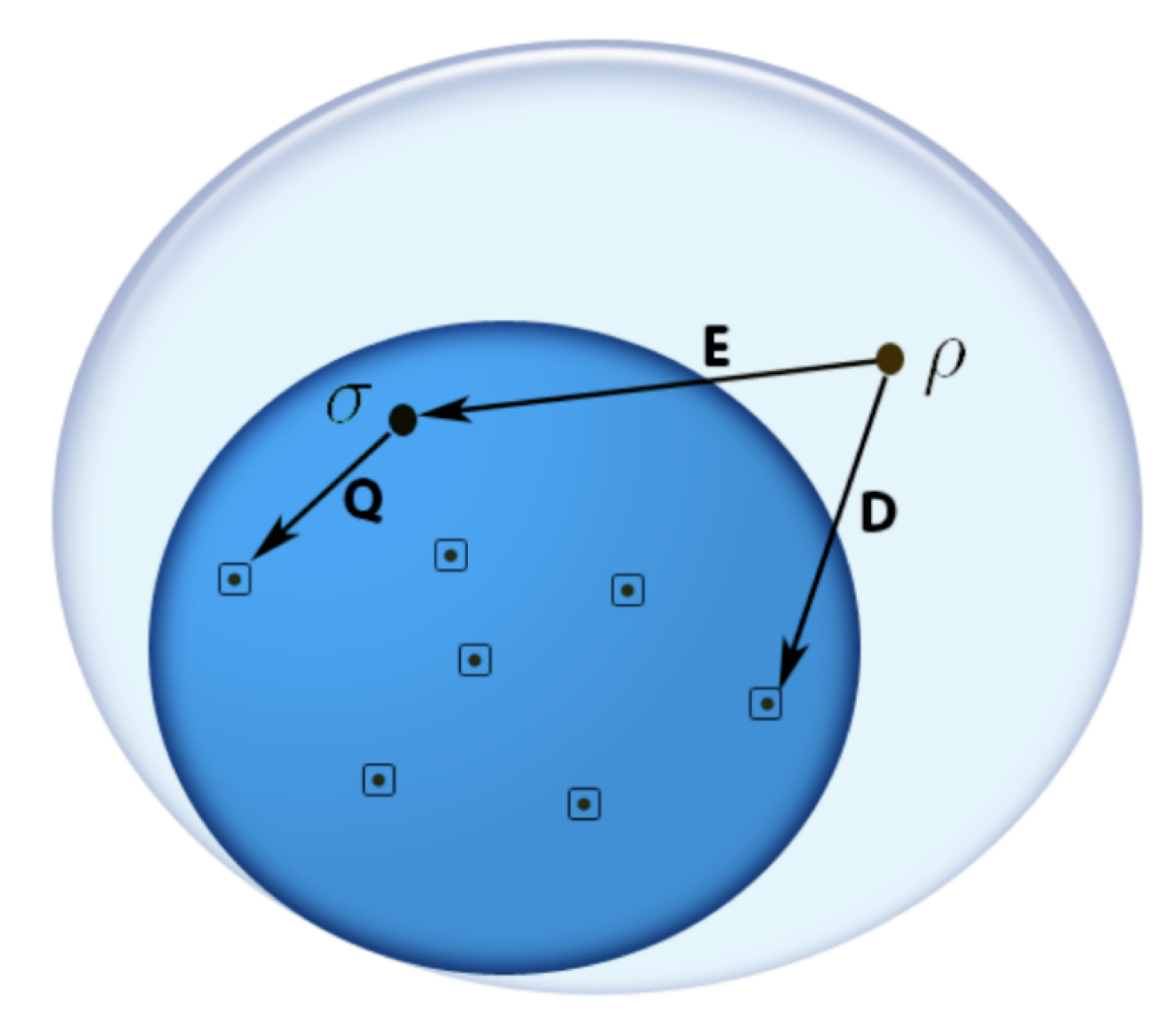}
\label{Fig:UnifiedView-B}
\caption{From \cite{Modi11}. The large ball represents the set of all quantum
states and the inside ball the set of separable states, which is convex. The squares
represents the set of classical states and the point the set of product states, neither
of them are convex. Note that our representation is just a sketch; actually the set of zero discord states, the classical ones, has null measure.}
\end{figure}

\section{Monogamy of quantum correlations}

One of the most important properties of entanglement is the fact that it
can not be freely shared between many parties. If two systems, $A$ and B, for
example, are maximally entangled, neither of them can be entangled with
a third system C. This comes from the fact that for $A$ and $B$ two be maximally entangled they should be in a pure state. On the other hand, if any of them is entangled it C, it should have non-zero entropy and therefore
be in a mixed state. But how about when $A$ and $B$ are not maximally entangled?n
In this situation $A$ and $B$ can be entangled with system $C$, but there are limits in the amount of entanglement they can share. These relations are named monogamy inequalities. The first and most well-know one was given by Cofmman,
Kundu and Wootters in 2000 \cite{Coffman00}, the CKW inequality,
\beq
C^2_{A|BC} \geq C^2_{A|B} + C^2_{A|C},
\eeq
with $C^2_{A|BC}$ representing the squared concurrence between $A$ and BC. So we
can see that given the amount of entanglement between $A$ and BC, the
amount of entanglement that $A$ and $B$ can share with C is restricted.
And if C increases its entanglement with $A$ or $B$, it has to decrease
the entanglement with the other. This relation can also be used
to define a measure of genuine tripartite entanglement as the
difference: $C^2_{A|BC} - C^2_{A|B} + C^2_{A|C} $. This measure
is usually named tangle or three tangle. It gives the intuition that
the entanglement between $A$ and BC is composed of the sum of bipartite entanglement of $A$ with $B$ and of $A$ with $C$, plus a genuine tripartite entanglement.
The relation is also valid for $N$ qubits \cite{Osborne06}:
\beq
C^2_{A_1|A_2...A_N} \geq C^2_{A_1|A_2} + C^2_{A_1|A_3} + ... +
C^2_{A_1|A_N}  
\eeq

Despite its appeal, the monogamy inequality above is not universal: not
true for all measures of entanglement. In particular it is not true for the entanglement of formation but only for its square. And when it is valid for qubits it usually breaks down in higher dimensions, with the only known exception being the squashed entanglement. And in fact, recently it has been shown that an important
class of entanglement measures may not obey a general monogamy relation for
arbitrary dimension \cite{Lancien16}.

Given the importance of monogamy relations, a natural question is if such a property is also true for other measures of quantum correlation. It was first noted that discord itself did not obey the CKW inequality for	three qubits; it was violated even for the W state. This could lead one to
say that discord is not monogamous. But we should be careful, as not
obeying the specific CKW inequality does not mean
it can be freely shared. It may obey other inequalities. And actually
not even the concurrence or the entanglement of formation obey
the CKW inequality, but only their squares. And it was later realized that
the square of discord does obey the CKW inequality\cite{Bai13}, being in this sense as monogamous as entanglement. In fact there is a stronger
relation between entanglement of formation and discord for three qubits.
This comes from an important monogamy relation between entanglement and
other correlations, obtained by Koashi and Winter in 2004 \cite{Koashi04}:
\beq
E_F(AB)+J_A(AC) \leq S(A),
\eeq
with $E_F$ being the entanglement of formation, $J_A(AC)$ the classical
correlation between $A$ and $C$ with measure in $A$ and $S(A)$ the usual von Neumann entropy
of A. The equality holds if $\rho_{ABC}$ is a pure state. By adding
the mutual information between $A$ and C on both sides we have that
for pure states
\beq
E_F(AB) = D_A(AC) + S(A|C),
\eeq
with $S(A|C)=S(AC)-S(C)$ the unmeasured conditional entropy; a
formula which can be used to obtain the discord. And as for pure
states $S(AC)=S(B)$ and $S(A|C)=S(B)-S(C)$ we have
\beq
D_A(AC) = S(C) - S(B) + E_F(AB)
\eeq
And with some further manipulation it is possible to show that \cite{Fanchini11}
\beq
D_A(AB)+D_A(AC)=E_F(AB)+E_F(AC)
\eeq
which is a monogamy relation between the sum of bipartite Discord and 
bipartite entanglement of formation in a three qubit system. It can also
be seen as a conservation law between bipartite discord and entanglement.
This also shows a relationship between the CKW inequality for discord 
and entanglement of formation. Actually, since for pure states discord
and entanglement are equivalent, so are the two inequalities \cite{Giorgi11}. But this is true only for pure states, with the distributed discord exceeding the distributed entanglement for mixed states. This same relationship is used to show that the squared discord does obey the CKW inequality.

The general question of which measure of quantum correlation obeys
the CKW inequality for general three-qudit states was addressed in \cite{Streltsov12}. It was shown that all measures of quantum
correlation beyond entanglement, which are nonzero on at least
one separable state, and obey some basic properties of a bona
fide measure are not monogamous in general. So there
is no good measure of quantum correlation which obey the CKW inequality 
for all states in any dimension. This can be true only for some
class of states or for some specific dimension, as the squared
Discord. We should also mention that the Hilbert-Schmidt geometric
discord also obeys the CKW inequality for three-qubit pure states.
In \cite{Streltsov12} is is also shown that any bona fide measure
of quantum correlation which obeys the CKW inequality can
not increase under local operations. And it is known that in general
quantum correlation can increase under local operations, which
can thus be connected with the lack of monogamy for such measures.
Recently it has also been shown that an important class of entanglement
measures can not satisfy an CKW type of inequality independent of
the dimension \cite{Lancien16}.

Another type of monogamy inequality was proposed in \cite{Braga12},
by replacing the bipartite quantum correlation between $A$ and BC
by a multiparite measure between $ABC$. More specifically, the following
inequality was proven
\beq
D(A:B:C) \geq D(A:B) + D(A:C)
\eeq
with $D(A:B:C)$ being the global quantum discord. However the inequality
can only be proven for quantum states whose conditional mutual information does not increase under measurement.

\section*{Conclusions}

In sum, it is now clear that entanglement is just one of
many interesting and intriguing characteristics of quantum
mechanics. This opens the possibility of new phenomena and applications
based not on entanglement but in these other forms of correlations.
And even tough these measures are not strictly correlation measures, they
may have some operational meaning, as already showed in some particular
situations, but still a question being explored. And the characterization of discord-like measures in condensed matter systems and dissipative system has also been a very active field; see \cite{Modi12} for more details of such works.

But as we mentioned before there are many possibilities to quantify
the quantumness of correlations. And many of them give qualitative different 
behaviors when characterizing physical systems, or just ordering the quantum
states by degree of quantumness. Even the choice of distance in geometric
measures can give such distinct behavior. This is actually not very surprising, given that it is a known result that different entanglement measures may also present different behaviors and induce different ordering in the state space.
But, while entanglement theory is well developed, although with still important
open questions, the discord and related measures have just started to be explored.


\vspace{1cm}

\noindent {\Large{\bf{Acknowledgment}}}
  
I would like to thanks Marcelo Sarandy for many discussion about discord and quantum correlation and Ernesto Galvão for carefully reading the first draft. I also acknowledge financial support from the Brazilian agencies CNPq, CAPES, FAPERJ, and the Brazilian National Institute of Science and Technology for Quantum Information (INCT-IQ).

\end{document}